\documentstyle[twocolumn,prb,aps]{revtex}

\begin{document}
\draft
\input{psfig}
\preprint{ }
% FOR TWO COLUMN ACTIVATE THE LINE BELOW 
\twocolumn[\hsize\textwidth\columnwidth\hsize\csname @twocolumnfalse\endcsname

\title{High Pressure Thermoelasticity of Body-centered Cubic Tantalum}

\author{O. G\"{u}lseren$^{(1,2,3)}$ and R.E. Cohen$^{(3)}$}
\address{$^{(1)}$ NIST Center for Neutron Research,
National Institute of Standards and Technology, Gaithersburg, MD 20899}
\address{$^{(2)}$ Department of Materials Science and Engineering,
University of Pennsylvania, Philadelphia, PA 19104}
\address{$^{(3)}$
Geophysical Laboratory and Center for High Pressure Research,
Carnegie Institution of Washington,
5251 Broad Branch Road, NW, Washington, DC 20015.}

\date{\today}
\maketitle

\begin{abstract}
We have investigated the thermoelasticity of body-centered cubic (bcc)
tantalum from first principles by using the linearized augmented plane
wave (LAPW) and mixed--basis pseudopotential methods for pressures up
to 400~GPa and temperatures up to 10000~K.
Electronic excitation contributions to the free energy were included
from the band structures, and phonon contributions were included using
the particle-in-a-cell (PIC) model. The computed elastic constants agree
well with available ultrasonic and diamond anvil cell data at low
pressures, and shock data at high pressures. The shear modulus $c_{44}$
and the anisotropy change behavior with increasing pressure around
150~GPa because of an electronic topological transition. We find that
the main contribution of temperature to the elastic constants is from
the thermal expansivity. The PIC model in conjunction with fast
self-consistent techniques is shown to be a tractable approach to
studying thermoelasticity.
\end{abstract}

%\vspace{2cm}

%62.20.Dc Elasticity, elastic constants
%62.20.-x Mechanical properties of solids 
%71.20.Be Transition metals and alloys
%46.25.Hf Thermoelasticity and electromagnetic elasticity
%46.25.-y Static elasticity
%
%05.70.Ce Thermodynamic functions and equations of state
%64.30.+t Equations of state of specific substances
%65.50.+m Thermodynamic properties and entropy
%65.70.+y Thermal expansion and density changes; thermomechanical effects

]

Single crystal elastic constants of solids at high pressures and
temperatures are essential in order to predict and understand material
response, strength, mechanical stability, and phase transitions.
We have studied the high pressure and temperature elastic constants of 
body-centered cubic (bcc) tantalum, a group V transition metal, from
{\sf first principles}. Because of its high structural mechanical, thermal
and chemical stability, Ta is a useful high pressure standard.\cite{cynn}
Ta has a very high melting temperature, 3269~K at ambient pressure.
Bcc Ta is stable to 174~GPa, according to diamond-anvil-cell
experiments.\cite{cynn} Shock compression experiments~\cite{hug1} show no
transition other than melting (at around 300~GPa). Its stability makes Ta
an ideal material for understanding the generic behavior of transition
metals under compression, without the complication of phase transitions.
Recently, its static properties were studied by full-potential LMTO
calculations~\cite{soderlind} and the thermal equation of state was
reported.\cite{tacohen}

The three elastic constants, $c_{11}$, $c_{12}$ and $c_{44}$, completely
describe the elastic behaviour of a cubic crystal. A more convenient set
for computations are $c_{44}$ and two linear combinations, $K$ and $c_s$.
The bulk modulus, 
\begin{equation}
K=(c_{11}+2c_{12})/3,
\end{equation}
is the resistance to deformation by an uniform hydrostatic pressure;
the shear constant,
\begin{equation}
c_s=(c_{11}-c_{12})/2,
\end{equation}
is the resistance to shear deformation across the (110) plane in the
$[1\bar{1}0]$ direction, and $c_{44}$ is the resistance to shear
deformation across the (100) plane in the $[010]$ direction.
The bulk modulus $K$ was determined from the equation of
state,\cite{tacohen} using the Vinet equation.\cite{vinet,ron1}
We obtained the shear moduli by straining the bcc lattice at fixed volumes
using volume conserving tetragonal and orthorhombic strains for $c_s$ and
$c_{44}$ respectively, and computing the free energy as a function of
strain. $c_s$ was obtained by applying the following isochoric strain
\begin{equation}
\epsilon = \left( \begin{array}{ccc}
\delta & 0 & 0 \\
0 & \delta & 0 \\
0 & 0 & (1+\delta)^{-2}-1
\end{array}
\right) ,
\end{equation}
where $\delta$ is the magnitude of the strain.
Then the strain energy is
\begin{equation}
F(\delta) = F(0) + 6 c_s V \delta^2 + O(\delta^3),
\end{equation}
where $F(0)$ is the free energy of the unstrained system and $V$ is its
volume. Similarly, $c_{44}$ was calculated from the following strain
\begin{equation}
\epsilon = \left( \begin{array}{ccc}
0 & \delta & 0 \\
\delta & 0 & 0 \\
0 & 0 & \delta^2/(1-\delta^2)
\end{array}
\right)
\end{equation}
with the corresponding strain energy
\begin{equation}
F(\delta) = F(0) + 2 c_{44} V \delta^2 + O(\delta^4).
\end{equation}
The quadratic coefficients of strain energy gives the elastic
constants. First order terms due to the initial stress
(hydrostatic pressure)~\cite{barron} were eliminated by applying isochoric
strains. Then, the elastic constants $c_{11}$ and $c_{12}$ were obtained from
$c_s$ and $K$.

We assume that the Helmholtz free energy of the system can be separated
as~\cite{fecons,wass,iron}:
\begin{equation}
F(V,T) = E_{0}(V) + F_{el}(V,T) + F_{vib}(V,T)
\end{equation}
where $E_{0}(V)$ is the static zero temperature energy,
$F_{el}(V,T)$ is the electronic contribution, and
$F_{vib}(V,T)$ is the vibrational contribution to the free energy.
Our computational procedure is based on density functional theory (DFT)
generalized to finite temperatures by the Mermin theorem.\cite{temdft}
The charge density is temperature dependent through occupation numbers
according to the Fermi-Dirac distribution, giving the electronic
entropy from
\begin{equation}
S_{el}=\sum{f_i ln f_i + (1-f_i) ln (1-f_i)}
\end{equation}
where $f_i=f_i(E-E_F,T))$ is the Fermi occupation at $T$ for each
state $i$. The variations of $f_i$ with temperature were included from
the self-consistent band structures calculated at an electronic
temperature of 2000~K varying according to the Fermi-Dirac distribution.

The electronic excitations, both the static energy and the electronic
contribution to the free energy, were computed by using the full potential
linearized augmented plane wave (LAPW) method.\cite{weikrakauer,singh}
The $5p$,$4f$,$5d$ and $6s$ states were treated as band states, and the
deeper states were treated as soft core electrons. The generalized gradient
approximation (GGA)\cite{gga-pbe} was used for the exchange-correlation
potential. The convergence of strain energies with respect to the Brillouin
zone integration was carefully checked by repeating the calculations for
16x16x16 and 24x24x24 meshes at V=16.82~\AA$^3${} and we found at most 2~GPa
(3 \%) difference both for $c_{s}$ and $c_{44}$. Hence, we used 16x16x16
special k-points meshes~\cite{monpack} in the full Brillouin zone giving
344 and 612 k-points within IBZ of tetragonal and orthorhombic lattice
respectively. The convergence parameter $R K_{max}$ was 9 giving about
1800 planewaves and 200 basis functions per atom at zero pressure.

The vibrational free-energy was obtained within the particle-in-a-cell
model (PIC)~\cite{cellmodel} by using an accurate pseudopotential
mixed-basis total energy method~\cite{mix-oguz} which is computationally
more efficient than the LAPW calculations. In PIC, an atom is displaced in
its Wigner-Seitz cell in the potential field of all the other atoms fixed
at their equilibrium positions, i.e. the ideal, static lattice except for
the wanderer atom. The partition function, and hence the free energy is
calculated from this potential energy surface via an integral over the
position of a single atom inside the Wigner-Seitz cell. The PIC model is
essentially an anharmonic Einstein model, and the 3N dimensional partition
function is reduced to a simple 3D integral.\cite{tacohen,wass}
The advantage of the cell model over lattice dynamics based on the
quasiharmonic approximation is that anharmonic contributions from the
potential-energy of the system have been included exactly without a
perturbation expansion. On the other hand, since we used the classical
partition function, and the interatomic correlations between the motions of
different atoms is ignored, it is only valid at high temperatures above the
Debye temperature (245 K in Ta.\cite{grigoriev}) Since the vacancy formation
energy is very high in Ta,\cite{sonali} spontaneous formation of defects is
only important after the melting temperature.

\begin{figure}
\centerline{\psfig{figure=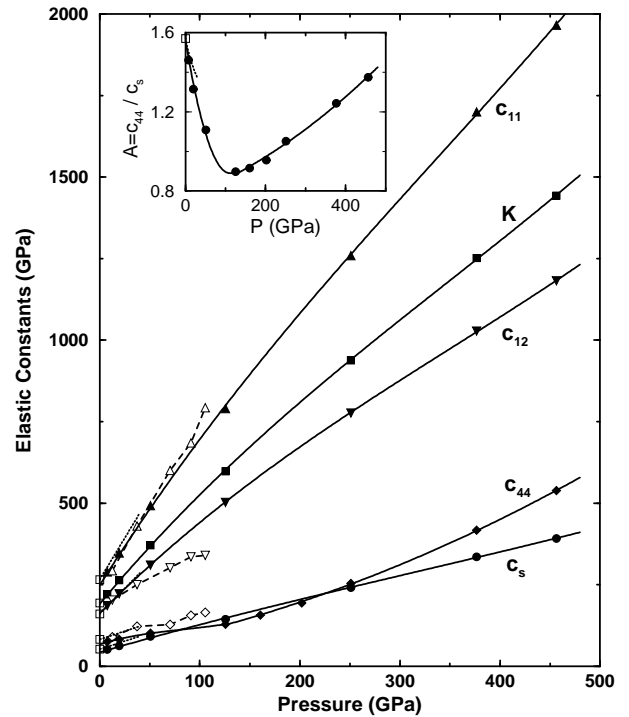,angle=0,width=82mm}}
%\vspace*{1mm}
\caption{\small Static elastic constants of Ta as a function of pressure.
Open squares are ultrasonic experimental data of Katahara 
{\it et al}\protect\cite{katahara1,katahara2}, and the dotted lines
show the initial slopes. Open symbols are SAX data of Cynn and
Yoo\protect\cite{cynnecons}.
The anisotropy ratio is shown in the inset.}
\label{fig:econst0K}
\end{figure}

For the PIC computations, a supercell with 54 atoms was used. The
pseudopotential mixed-basis calculations were carried out on this 54 atoms
supercell using LDA~\cite{ldazun} for exchange-correlations effects
and 2x2x2 k-point mesh resulting 4 special $\vec{k}$ points for BZ
integrations. A semi-relativistic, nonlocal and norm-conserving
Troullier-Martins~\cite{tromartins} pseudopotential (with associated
pseudo-atomic orbitals) with non-linear core corrections~\cite{nonlinearcore}
was used for the Ta atoms as described in detail in our previous study
of thermal equation of state of Ta.\cite{tacohen} After checking the
energy convergence, 550~eV and 60~eV are used as planewave energy cutoffs
for the expansion of the pseudo-atomic orbitals as well as FFT grid and
low-energy plane waves for additional degrees of freedom in basis set
respectively. The canonical partition function was computed from the
potential energy surface as a function of displacements of wanderer atom
along special symmetry directions.\cite{wass,spdir} We used 2 and 4 special
directions for tetragonal and orthorhombic distortions, respectively,
which integrates exactly up to $l=6$ lattice harmonics.\cite{spdir} The
potential energy was calculated at 4-6 different displacements along each
of these special directions, and was fit to an even polynomials up to order
six. Details of the all computational parameters were described
previously.\cite{tacohen}

The static elastic constants as functions of pressure are presented in
Fig.~\ref{fig:econst0K} and Table~\ref{table:econs}. The zero pressure
values and initial slopes are
in good agreement with the ultrasonic experimental data of Katahara
{\it et al.}\cite{katahara1,katahara2} Similarly, comparison with recent SAX
(stress/angle-resolved x-ray diffraction) experimental data~\cite{cynnecons}
up to 105~GPa shows good agreement for $c_{11}$ and $c_{44}$. Likewise,
$c_{12}$ agrees well at low pressures, but deviates with increasing pressure.
This may be due to the assumed isostress condition for experimental
data analysis for all pressures, or due to the large uncertainty on measured
deviatoric stress at high pressures. Note that, the initial slope of
ultrasonic data agrees very well with our calculated $c_{12}$.
The anisotropy ratio, $A=c_{44}/c_{s}$ (inset Fig.~\ref{fig:econst0K})
first decreases from 1.57 to 0.9 with increasing pressure, and then its
slope reverses and it increases with increasing pressure. This is due to
the changes in $c_{44}$. The change in behavior of $c_{44}$ around 150 GPa is
due to the electronic transition evident in the equation of
state.\cite{tacohen} This indicates that elastic constants can be much more
sensitive to changes in the Fermi surface than the equation of state, where
the electronic transition was not apparent without examining small residuals
in the equation of state fit.

\begin{figure}
\centerline{\psfig{figure=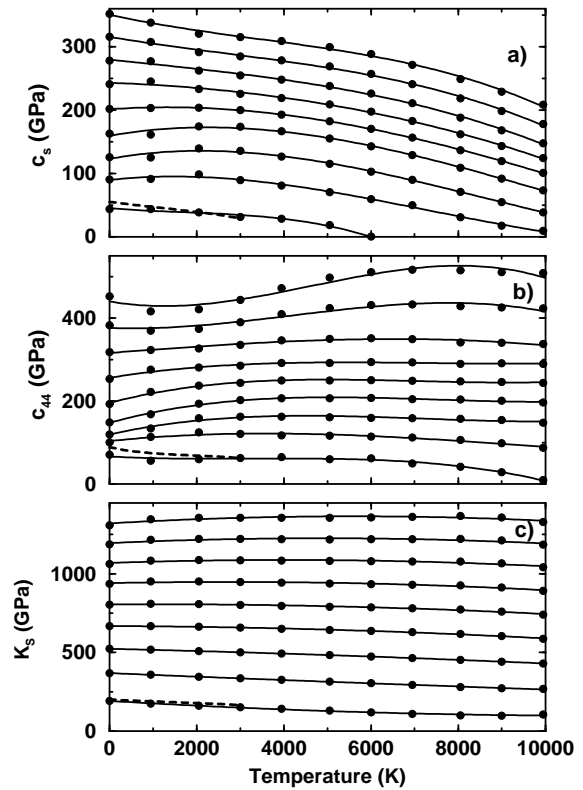,angle=0,width=80mm}}
%\vspace*{3mm}
\caption{\small The elastic constants of bcc Ta as a function of
temperature at different pressure from 0~GPa (lowest curve)
to 400~GPa (uppermost curve) with 50~GPa interval.
a) shear modulus $c_{s}$, b) shear modulus
$c_{44}$ and c) adiabatic bulk modulus $K_S$.
Dotted lines are the experimental data from Walker
et al.~\protect\cite{walker}.}
\label{fig:econsther}
\end{figure}

The elastic constants of Ta as functions of temperature at various
pressures are presented in Fig.~\ref{fig:econsther} and
Table~\ref{table:econsther}.
In order to compare with experimental data, the computed isothermal elastic
constants ($c_{ij}^{T}$) are converted to adiabatic constants ($c_{ij}^{S}$)
according to~\cite{davies}
\begin{equation}
c_{ij}^{S} =  c_{ij}^{T} + \frac{T}{\rho C_V} \lambda_i \lambda_j
\label{eq:isoadb}
\end{equation}
where $\lambda_i = \sum_k\alpha_k c_{ik}^{T}$, $\alpha_k$ is the linear
thermal expansion tensor, $C_V$ is the specific heat and $\rho$ is the
density. For cubic crystals, eq.~\ref{eq:isoadb} simplifies to
\begin{eqnarray}
c_{11}^{S} & = & c_{11}^{T} + \Delta \\
c_{12}^{S} & = & c_{12}^{T} + \Delta
\end{eqnarray}
where
\begin{equation}
\Delta = T (\alpha K_T)^2/(\rho C_V) = \rho C_V T \gamma^2 = 
T \alpha K_T \gamma
\end{equation}

\begin{table}[h]
\begin{tabular}{r r r r r r r}
%\hline
V (\AA$^3$) & Pressure & K & $c_{44}$ & $c_s$ & $c_{11}$ & $c_{12}$ \\
\hline
18.39 &  -0.76 &  187.89 &  66.30 &  42.12 &  244.05 &  159.82 \\
17.66 &   7.56 &  220.35 &  75.13 &  51.41 &  288.90 &  186.08 \\
16.82 &  19.35 &  263.82 &  82.73 &  62.89 &  347.68 &  221.90 \\
15.22 &  50.88 &  371.07 & 101.15 &  91.30 &  492.79 &  310.20 \\
13.01 & 125.70 &  598.02 & 129.18 & 143.78 &  789.73 &  502.17 \\
12.43 & 160.63 &  696.44 & 156.64 &        &         &         \\
11.67 & 202.01 &  808.93 & 194.45 &        &         &         \\
11.03 & 250.90 &  937.44 & 253.95 & 241.48 & 1259.41 &  776.46 \\
 9.83 & 376.54 & 1251.81 & 417.35 & 335.65 & 1699.34 & 1028.04 \\
 9.26 & 456.48 & 1443.14 & 538.35 & 391.79 & 1965.52 & 1181.94 \\
%\hline
\end{tabular}
\vspace*{0.1cm}
\caption{The static elastic constants for bcc Tantalum. All elastic
constants as well as pressure units are GPa.}
\label{table:econs}
\end{table}
with $\alpha$ is the thermal expansion coefficient, $\gamma$ is the
Gr\"{u}neisen parameter, and $K_T$ is the isothermal bulk modulus. The
thermodynamic parameters were computed self-consistently from the thermal
equation of state.\cite{tacohen} The correction is zero for $c_{44}$ and
$c_{s}$. $\Delta$ increases with temperature but decreases with pressure;
at 3000~K it decreases from 5~\% to 1~\% for pressures 50~GPa to 400~GPa
for bulk modulus $K_T$, and at 10000~K $\Delta$ is 29~\% and 3~\% for
same pressures.

\begin{table}[h]
\begin{tabular}{c r r r r r r r r}
%\hline
  T (K) & $c_s$ & $c_{44}$ & $c_{11}^T$ & $c_{11}^S$ &
$c_{12}^T$ & $c_{12}^S$ & $K_T$ & $K_S$ \\
\hline
   0 & 44.05 & 70.26 & 249.68 & 249.68 & 161.59 & 161.59 & 190.95 & 190.95 \\
 947 & 43.58 & 56.24 & 221.71 & 233.41 & 134.55 & 146.27 & 163.61 & 175.33 \\
2053 & 38.61 & 59.39 & 189.66 & 211.95 & 112.44 & 134.72 & 138.18 & 160.46 \\
3000 & 31.57 & 62.48 & 162.55 & 192.74 &  99.41 & 129.60 & 120.46 & 150.65 \\
3947 & 28.72 & 64.47 & 142.15 & 179.77 &  84.70 & 122.32 & 103.85 & 141.47 \\
5052 & 18.34 & 59.70 & 107.55 & 154.15 &  70.87 & 117.47 &  83.10 & 129.70 \\
6000 &  0.31 & 62.05 &  64.91 & 120.03 &  64.29 & 119.40 &  64.50 & 119.61 \\
%\hline
\end{tabular}
\vspace*{0.2cm}
\caption{\small The elastic constants for bcc Tantalum at various
temperatures. All elastic constants units are GPa.}
\label{table:econsther}
\end{table}

The shear moduli $c_s$ and $c_{44}$ and adiabatic bulk modulus $K_S$ agree
well with the ultrasonic experimental data~\cite{walker} up to 3000~K at
zero pressure (Fig.~\ref{fig:econsther}). We find that all three moduli are
primarily functions of volume, and thermal effects at constant volume are
quite small except at the highest pressures. There is some softening of $c_s$
with increasing temperature for all pressures. $c_{44}$, shows a slight
softening at the zero pressure with increasing temperature, but they are
rather flat for other pressures except than very high pressures.
The adiabatic bulk modulus $K_S$ also softens slightly with temperature at
low pressures but becomes flat with increasing pressure.

\begin{figure}
\centerline{\psfig{figure=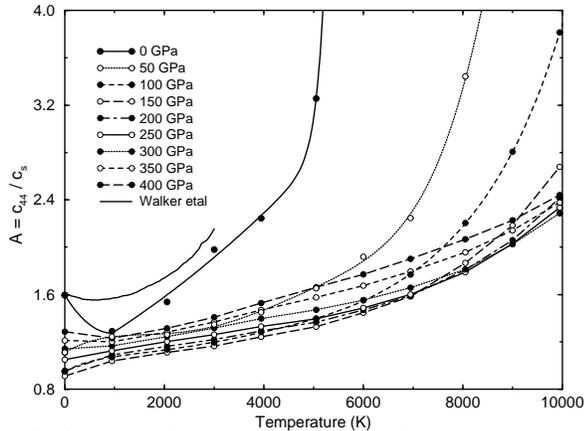,angle=-90,width=80mm}}
%\vspace*{1cm}
\caption{\small The anisotropy ratio of elastic constants of Ta as a
function of temperature at different pressures from 0~GPa to 400~GPa with
50~GPa interval.}
\label{fig:anist}
\end{figure}

The anisotropy ratio, $A=c_{44}/c_s$, is presented as a function of
temperature for various pressures at Fig.~\ref{fig:anist}.
$A$ increases with increasing temperature at all pressures, but less
drastically at high pressures. At lower pressures, this increase is
divergent after certain temperature, since the softening of $c_s$ is
large enough and it approaches zero. The reversal of the slope of $A$
with pressure shifts to higher pressures with increasing temperature due
to thermal expansivity, and occurs at a fixed volume.

\begin{figure}
\centerline{\psfig{figure=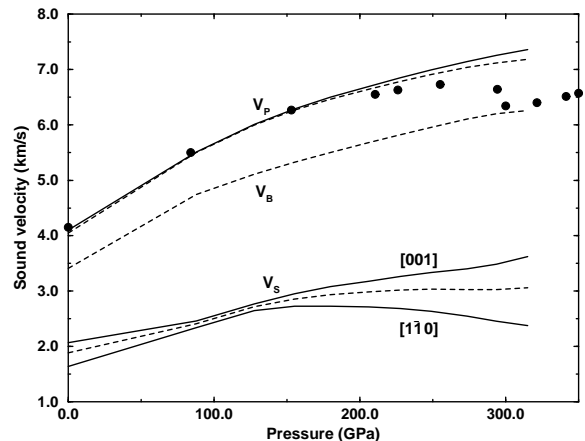,angle=-90,width=80mm}}
%\vspace*{1cm}
\caption{\small Sound velocities of Ta along the Hugoniot calculated 
from elastic constants. Solid lines are the longitudinal and two
transverse sound velocities in $[110]$ direction from single crystal
elastic constants. The polarization of the shear waves are given
in brackets. The isotropic aggregate velocities are shown by dashed
lines. $v_P$, $v_B$, and $v_S$ are the compressional, bulk, and shear
sound velocities.
Filled dots are the shock data from
Brown and Shaner~\protect\cite{hug2}.}
\label{fig:sound}
\end{figure}

Sound velocities are related to the elastic constants by the Christoffel
equation~\cite{christoffel}
\begin{equation}
(c_{ijkl} n_j n_k - \rho v^2 \delta_{ij}) u_i = 0,
\end{equation}
where $c_{ijkl}$ is the elastic constants tensor, $\vec{n}$ is the
propagation direction, $\vec{u}$ is the polarization vector and
$v$ is the velocity. Our elastic constants are those appropriate for the
equations of motion under hydrostatic reference stress.\cite{barron}
For $[110]$ wave propagation direction in a cubic lattice, the
longitudinal mode is
\begin{equation}
\rho v^2 = (c_{11} + c_{12} + 2c_{44})/2
\end{equation}
and two transverse modes are
\begin{equation}
\rho v^2 = c_{44}
\end{equation}
and
\begin{equation}
\rho v^2 = (c_{11} - c_{12})/2 = c_s
\end{equation}
polarized along $[001]$ and $[1\bar{1}0]$ direction respectively.
For polycrystalline sample, the average isotropic shear modulus $G$ can
be determined from single crystal elastic constants according to the
Voigt-Reuss-Hill scheme~\cite{vrhaveg}, and the isotropically averaged
aggregate velocities are given by
\begin{eqnarray}
v_P & = & \left( (K+4/3G)/\rho\right)^{1/2} \\
v_S & = & (G/\rho)^{1/2} \\
v_B & = & (K/\rho)^{1/2}
\end{eqnarray}
where $v_P$, $v_S$, and $v_B$ are the compressional, shear and bulk
sound velocities. The sound velocities of Ta along the Hugoniot calculated
from elastic constants are shown in Fig.~\ref{fig:sound}, and are
compared with the shock sound velocity data from Brown {\it et al.}\cite{hug2}
As seen in Fig.~\ref{fig:sound}, there is excellent agreement with
shock data. The calculated compressional velocity $v_P$ agrees very well
with experimental data up to 200~GPa, and then after 300~GPa the bulk
velocity $v_B$ matches the data well. This is because the shocked
solid melts around 300~GPa, so the liquid velocity might be represented
by $V_B$. The deviation between 200~GPa and 300~GPa is probably due to
premelting effects.

In conclusion, the elasticity of bcc Ta is investigated from first
principles for pressures up to 400~GPa and temperatures up to 10000~K.
The calculated static elastic constants are in good agreement with
available ultrasonic and SAX experimental data. The shear
modulus $c_{44}$ and the anisotropy ratio $A$ change behaviour with
increasing pressure around 150~GPa. Although, the shear modulus $c_s$
softens with increasing temperature at all pressures, $c_{44}$ and
$K_S$ soften with temperature at low pressures but then they are rather
flat at higher pressures. The main effect of temperature for the
thermoelasticity of Ta is due to thermal expansivity. The calculated
sound velocities along the Hugoniot shows an excellent agreement with
shock-wave experimental data.

{\bf Acknowledgments}
We thank G. Steinle-Neumann, L. Stixrude, and E.  Wasserman for
helpful discussions. This work was supported by DOE ASCI/ASAP
subcontract B341492 to Caltech DOE W-7405-ENG-48.
Computations were performed on the Cray SV1 at the Geophysical Laboratory,
supported by NSF grant EAR-9975753 and the W.\ M.\ Keck Foundation.

%\clearpage

\end{document}